\documentclass[useAMS,usenatbib]{mn2e}

\usepackage {graphicx}
\usepackage {times}

\title[Abundance diagnosis of E+A  (post-starburst) galaxies]{Abundance diagnosis of E+A (post-starburst) galaxies}
\author[T. Goto]{Tomotsugu Goto
 \thanks{E-mail:tomo@ir.isas.jaxa.jp} 
\\
 Institute of Space and Astronautical Science,  Japan Aerospace Exploration Agency,
 3-1-1 Yoshinodai, Sagamihara, Kanagawa 229-8510, Japan}
\begin{document}

\date{\today; in original form 2006 November 1}

\pagerange{\pageref{firstpage}--\pageref{lastpage}} \pubyear{2006}

\maketitle

\label{firstpage}

\begin{abstract}
 E+A galaxies are characterized as a galaxy with strong Balmer
 absorption lines but without any [OII] or H$\alpha$ emission
 lines. The existence of strong Balmer absorption lines means that E+A
 galaxies have experienced a starburst within the last $<$1-1.5 gigayear. However, the
 lack of [OII] or H$\alpha$ emission lines indicates that E+A galaxies
 do not have any on-going star formation. Therefore, E+A galaxies are
 interpreted as a post-starburst galaxy. 
 Morphologically, E+A galaxies appear as early-type galaxies, implying E+A galaxies may be one of the progenitors of present-day elliptical galaxies.
 However, there remained other possibilities such as the dusty starburst scenario, where E+A galaxies have on-going star formation, but optical emission lines are invisible due to the heavy obscuration by dust. Therefore, an additional evidence of the post-starburst phenomena has been eagerly awaited.

 Using  one of the largest samples of 451 E+A galaxies carefully selected from the Sloan Digital Sky Survey Data Release 4,
 here we show the abundance diagnosis of E+A galaxies using Mg and Fe lines.
Our findings are as follows :(i) E+A galaxies has enhanced $\alpha$-element abundance ratio compared to the star-forming galaxies with similar Balmer absorption strength. Since the truncation of strong starburst is required to enhance the alpha element ratio, this is an additional evidence that E+A galaxies are in the post-starburst phase; (ii) the metallicity and $\alpha$-element abundance of E+A galaxies are consistent with those of elliptical galaxies, suggesting that E+A galaxies could be one of the progenitors of present-day elliptical galaxies in terms of chemical abundances.

\end{abstract}

\begin{keywords}
galaxies: evolution, galaxies:interactions, galaxies:starburst, galaxies:peculiar, galaxies:formation
\end{keywords}

\section{Introduction}

\citet{1983ApJ...270....7D,1992ApJS...78....1D} found galaxies with mysterious spectra while
 investigating high redshift cluster galaxies.
 The galaxies had strong Balmer absorption lines with no
 emission in [OII]. These galaxies are called  ``E+A''
 galaxies since their spectra looked like a superposition of that of
 elliptical galaxies (Mg$_{5175}$, Fe$_{5270}$ and Ca$_{3934,3468}$
 absorption lines) and that of A-type stars (strong Balmer absorption lines).
  The existence of strong Balmer absorption lines shows 
 that these galaxies have experienced starburst recently (within a
 gigayear; Goto 2004b).  However, these galaxies do not show any sign of on-going star
 formation as non-detection in the [OII] emission line
 indicates.  
   Therefore, E+A galaxies have been interpreted as a post-starburst galaxy,
 that is,  a galaxy which truncated starburst suddenly \citep{1983ApJ...270....7D,1992ApJS...78....1D,1987MNRAS.229..423C,1988MNRAS.230..249M,1990ApJ...350..585N,1991ApJ...381...33F,1996ApJ...471..694A}.  


 However, there remain other possibilities: 
   E+A galaxies could also  be explained as dusty-starburst galaxies. In this scenario, E+A galaxies are not post-starburst galaxies, but in reality star-forming galaxies whose emission lines are invisible in optical wavelengths due to the heavy obscuration by dust. 
  \citet{1999ApJ...525..609S}  performed a radio observation to investigate the star formation activity hidden by the dust, and detected 5 galaxies in radio (out of 8), which have strong Balmer absorption with no detection in [OII]. 
   \citet{1999AJ....118..633O} investigated the radio properties of galaxies in
  a rich cluster at $z\sim$0.25 (A2125) and found that optical line
  luminosities (e.g., H$\alpha$+[NII]) were often weaker than one would
  expect for the star formation rates (SFRs) implied by the radio emission.
    As a variant,  \citet{2000ApJ...529..157P} presented the selective dust extinction
   hypothesis, where dust extinction is dependent on stellar age since 
    youngest stars inhabit very dusty star-forming HII regions while older
    stars have had time to migrate out of such dusty regions. 
     If O, B-type stars in E+A galaxies are embedded in dusty regions and
    only A-type stars have long enough lifetimes ($\sim$ 1 gigayear) to move out from such regions,  
    this scenario can naturally explain the E+A phenomena.
   \citet{2006ApJ...649..163B} detected large amount of neutral hydrogen ($10^9 M_{\odot}$)in four nearby E+A galaxies, suggesting that some E+A galaxies are observed in the inactive phase of their star formation duty cycle. 
 Accordingly, an independent evidence of the post-starburst event in addition to the strong Balmer absorption has been awaited. 

 The abundance of  $\alpha$-element  can carry crucial information about the star formation time-scales.
$\alpha$-elements such as Magnesium\footnote{$\alpha$-element includes O, Mg, Si, S, Ca and T.} are formed in the explosion of type II supernovae which occur rapidly after a burst of star formation (time scales of $10^6-10^7$ yr) whereas the iron peak elements originate in Type Ia supernovae which lag behind by at least 1 gigayear \citep{1984ApJ...286..644N,1995ApJS..101..181W}. 
Therefore, $\alpha$-elements enhancement is determined by the duration of star formation, i.e., a super-solar $\alpha$-element to Fe ratio indicates that stars formed in an initial burst tailing up the Type II supernova abundance pattern and then star formation ceased. 
 If E+A galaxies are truly in the post-starburst phase as has been claimed before \citep[e.g.,][]{1999ApJ...518..576P}, E+A galaxies are expected to have an enhanced $\alpha$-element ratio, where Type II SNe produced $\alpha$-elements such as Mg, but the Type I SNe has not yet exploded to produce enough Fe. 
 In environments where star formation can continue for a longer period
 of time (e.g., the disc of our own galaxy), one  expects solar
 abundance ratio for the younger, more metal-rich stars, which indeed is
 the case \citep{1989PhDT.......149P,1992ApJ...398...69W,1993A&A...275..101E,1997ARA&A..35..503M,1997ApJS..111..203V}.
Note that it is true that there will be some type Ia SNe from the remaining star in E+A galaxies after they truncate the starburst to create some Fe. However, since the E+A galaxies have negligible amount of the remaining star formation activity, there is little chance for this newly created Fe to get into the stellar spectra.

 In the mean time, there has been a claim that E+A galaxies could be a progenitor of present day elliptical galaxies \citep{2004ApJ...602..190Q,2006ApJ...650..763H}.  
 It has been found that E+A galaxies have early-type morphological appearance \citep{2005MNRAS.357..937G}. Some of E+A galaxies have even dynamically disturbed appearance \citep{1991ApJ...381L...9O,2005MNRAS.359.1557Y,2005MNRAS.359.1421P,2006ApJ...650..763H}.
Recently, \citet{2005MNRAS.357..937G} has shown that E+A galaxies have more close companion galaxies than average galaxies, showing that the dynamical merger/interaction could be the physical origin of E+A galaxies. Since the dynamical computer simulations have shown that the dynamical merger can create an elliptical galaxy \citep{1978MNRAS.184..185W,1992ApJ...393..484B,2001ApJ...547L..17B,2001ApJ...558...42S}, E+A galaxies are a perfect candidate of the progenitor of present-day elliptical galaxies, although two orders of magnitude of difference in the number density needs to be carefully considered.
 If E+A galaxies are one of the progenitors, they should have similar metallicity/abundance ratios to those of present-day elliptical galaxies.

 One of the difficulties in investigating E+A galaxies has been their rarity: less than 1\% galaxies are in the E+A phase in the present Universe \citep{2003PASJ...55..771G}. 
 In this work, we overcome this problem by creating a large sample of 451 E+A galaxies carefully selected from the fourth public data release of the Sloan Digital Sky Survey \citep[SDSS;][]{2006ApJS..162...38A}, and investigate metallicity, abundance ratios and $\alpha$-element enhancement of E+A galaxies. The aim of this study is as follows:
(i) By investigating the metallicity and  $\alpha$-element enhancement of 451 E+A galaxies, we seek an independent evidence that E+A galaxies are in truly the post-starburst phase; (ii) By comparing the abundance ratio and the $\alpha$-element enhancement of E+A galaxies to those of luminous elliptical galaxies, we test if E+A galaxies can be a progenitor of present-day elliptical galaxies.


   Unless otherwise stated, we adopt the best-fit WMAP cosmology: $(h,\Omega_m,\Omega_L) = (0.71,0.27,0.73)$ \citep{2003ApJS..148....1B}.

\section{Sample}\label{data}

\subsection{Data}

 We have created a new sample of 451 E+A galaxies using the fourth public data release of the SDSS \citep{2006ApJS..162...38A}. This is the largest sample of spectroscopically selected E+A galaxies to date. The selection algorithm is 
essentially the same as that in \citet{2005MNRAS.357..937G}, which we briefly summarize below.

  We only use those objects classified as galaxies ({\tt type=3}, i.e., extended) with spectroscopically measured redshift of $z>0.01$ and the spectroscopic signal-to-noise $>$10 per pixel (in the continuum of the $r$-band wavelength range). 
 For these galaxies, we have measured H$\delta$,
  [OII] and H$\alpha$ equivalent widths (EWs) and obtained their errors
  using the flux summing  method described in \citet{2003PASJ...55..771G}. 
  For the H$\delta$ line, we only used the wider window of  4082$-$4122\AA\ to ease the comparison with models.
  Once the lines are measured, we have selected E+A galaxies as those with H$\delta$~EW~$>$~5.0\AA, and H$\alpha$ EW $>$ $-$3.0\AA, and [OII] EW $>$ $-$2.5\AA\footnote{Absorption lines have a positive sign throughout this paper.}. Although the criteria on emission lines allow small amount of emission in the E+A sample, they are relatively small amount in terms of the SFR.   We also exclude galaxies at $0.35<z<0.37$ from our sample because of the sky feature at 5577\AA\ .
 Our criteria are more strict than previous ones (e.g., H$\delta$ EW $>$ 4.0\AA\ and [OII]~EW~$> -$2.5\AA), suppressing possible contaminations from other populations of galaxies \citep{2004A&A...427..125G}.
 
We stress the advantage in using the H$\alpha$ line in selecting E+A galaxies. Previous samples of E+A galaxies were often selected based solely on [OII] emission and Balmer absorption lines either due to the high redshift of the samples or due to instrumental reasons. \citet{2003PASJ...55..771G} showed that such selections of E+A
 galaxies without information on H$\alpha$ line would suffer from 52\%
 of contamination from H$\alpha$ emitting galaxies, whose morphology and color are very different from that of E+A galaxies \citep{2003PASJ...55..771G}.
\citet{2004MNRAS.355..713B} selected E+A galaxies from the 2dF using only Balmer and [OII] lines to find that some E+A galaxies in their sample have the H$\alpha$ line in emission. 
Since H$\beta$ and H$\gamma$ absorption features are subject to the emission filling, \citet{2004MNRAS.355..713B} found that using three Balmer absorption lines (H$\delta$, H$\gamma$ and H$\beta$) can supress the contamination from the galaxies with detectable H$\alpha$ emission. 
 
 Among $\sim$210,000 galaxies which satisfy the redshift and S/N
  criteria with measurable [OII], H$\delta$ and H$\alpha$ lines, we have found 451 E+A galaxies, which is one of the largest sample of E+A galaxies to date. 
  The 451 E+A galaxies span a redshift range of  0.027 $\leq z \leq$ 0.394, within which H$\alpha$ line is securely covered by the SDSS spectrograph. Although this range covers about 2 gigayear of lookback time, majority of our E+A galaxies lie at $z\sim 0.1$. Therefore, we do not consider the evolutionary effect within the E+A sample in this paper.
It is important to note that the SDSS fiber spectrograph only samples the inner 3'' of the region in diameter.  Therefore, for the lowest redshift E+As, the 3''-fiber only samples the inner bulge region. And thus, readers are cautioned that it is possible for these E+A galaxies to have some remaining star formation activity in the outside of the fiber.  However, since these galaxies at least have a post-starburst phenomena at the centre of the galaxy, we still consider these galaxies as E+A galaxies in this work.   
In addition, since majority of our sample lie at $z\sim 0.1$, there is little variation in terms of the sampling region within our E+A galaxies \citep[see][for relevant description on the aperture effects]{2003MNRAS.346..601G}.

For a comparison purpose, we also select two different populations of galaxies: we select emission line galaxies with comparable Balmer line strength of $H\delta$ EW $>$ 5.0\AA together with both H$\alpha$ and [OII] in emission (H$\alpha$ EW$<$-3.0\AA and [OII] EW $<$-2.5\AA). The same redshift and signal-to-noise ratio cuts are applied to these galaxies.  AGNs are removed from the sample using the [SII]/H$\alpha$ vs [OIII]/H$\beta$ line ratios \citep[see][for more detail]{2005MNRAS.356L...6G}. Median star formation rate based on the extinction corrected H$\alpha$ flux is 9.3 M$_{\odot}$ yr$^{-1}$. For these galaxies, Balmer absorption lines suffer from significant amount of the emission filling, which we have corrected using the H$\alpha$/H$\beta$ ratio and the iteration procedure described in \citet{2003PASJ...55..771G}. There remained 15886 galaxies and we call these galaxies as star-forming galaxies hereafter. 

Another comparison sample is selected with H$\alpha$ EW $>$ $-$3.0\AA, and [OII] EW $>$ $-$2.5\AA (in addition to the same redshift and SN cuts), i.e., galaxies with no emission in H$\alpha$ nor [OII]. We call them as passive galaxies, and  there are $\sim$16500 such galaxies. Although we regard this as a representative of present-day elliptical galaxies, readers are cautioned that this passive galaxy sample includes lower mass systems than the so-called giant elliptical galaxies.

\subsection{Lick indices}\label{lick}

For these E+A, star-forming, and the rest of the galaxies, we measured Mgb, Fe5270, and Fe5330 Lick/Image dissection scanner (IDS) indices based on the definition of \citet{1994ApJS...94..687W,1997ApJS..111..377W}. The Lick system involves measuring line fluxes in narrow wavelength ranges relative to a pseudo-continuum found by linearly interpolating between two spectral averages on either side of the lines.  
In order to compare measured indices with theoretical model predictions and the literature, the measurements have to be carefully calibrated to the Lick/IDS system, i.e., we have to account for (i) the difference in the spectral resolution between Lick/IDS system and the SDSS spectra; (ii) the internal velocity broadening of the observed galaxies. 
 To achieve these, we convolved each spectrum with a wavelength-dependent Gaussian so that the broadened spectra together with the velocity broadening reproduce the Lick/IDS resolution presented in Fig.7 of \citet{1997ApJS..111..377W}. For galaxies without proper velocity dispersion measurement, we have assumed a median velocity dispersion of our sample (175 km s$^{-1}$).
 Strictly speaking, we further need to adjust the flux calibration to the Lick system using a defined set of Lick standard stars, which have not been observed with the SDSS telescope unfortunately. The difference in the flux calibration, however, brings much smaller effects than those of the resolution  and the velocity broadening.

\section{Results}\label{results}

\subsection{Mgb and Fe5270}

\begin{figure}
\begin{center}
\includegraphics[scale=0.6]{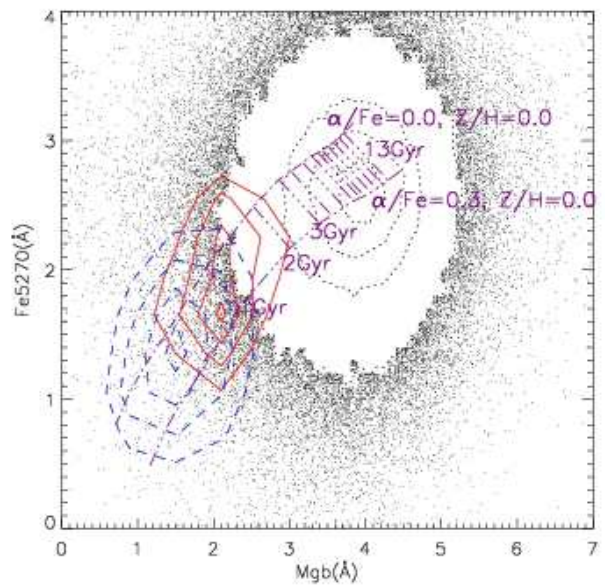}
\end{center}
\caption{
Equivalent width of Fe(5270\AA) is plotted against Mg$_b$. 
The dotted contours and the small dots show the distribution of passive
 galaxies (with no emission lines). The solid contours and the dashed
 contours show the distribution of E+A and the star-forming galaxies,
 respectively. The dash-dot lines are the models of
 \citet{2003MNRAS.339..897T} with $\alpha/Fe=0.3$ and 0.0 (where
 $Z/H=0.0$), where 1,2,3, and 13 gigayears of ages are labeled. 
 The countour levels are at  20,40,60 and 80 percentiles.
E+A galaxies have slightly larger metallicity than the star-forming galaxies, and have similar $\alpha$-abundance with elliptical galaxies.
}\label{fig:FeMg}
\end{figure}

In Fig \ref{fig:FeMg}, we plot equivalent width of Fe(5270\AA) against the Mg$_b$ index.
The dotted contours and the small dots show the distribution of passive galaxies in the SDSS spectroscopic sample (selected with H$\alpha$ EW $>$ $-$3.0\AA, and [OII] EW $>$ $-$2.5\AA, i.e., no H$\alpha$ nor [OII] in emission). 
 We overplot models  of \citet{2003MNRAS.339..897T}, which provide simple stellar population models with variable element abundance ratios. These models are free from the intrinsic $\alpha$/Fe bias that was imposed by the Milky Way template stars, and thus, the models reflect well-defined  $\alpha$/Fe ratios at all metallicities. 
 This treatment of non-solar abundance ratios is one of the most crucial elements in determining absolute age/metallicity estimates.
Although starburst galaxies generally have two distinct stellar populations, i.e., bursting population and the underlying old stellar population, according to the population synthesis models, if more than 5\% of the galaxy mass experience the burst, the galaxy spectrum is dominated by the bursting population  \citep[see ][ for the detail]{2006ApJ...642..152Y,2006AJ....131.2050Y}.
  The dash-dot lines are the models with $\alpha/Fe=0.3$ and 0.0 (where $Z/H=0.0$). Along the model lines, 1,2,3, and 13 gigayears of ages are labeled. 

 Since the late 1970s, evidence has been accumulating  that the magnesium-to-iron ratio seems to be larger in luminous elliptical galaxies when compared to solar-neighborhood stars \citep{1976ApJ...206..370O,1989PhDT.......149P,1992ApJ...398...69W,1993MNRAS.262..650D,1999PASP..111..919H,1999MNRAS.306..607J}. 
Compared to the $\alpha/Fe=0.0$ model, the distribution of all the passive galaxies has significant $\alpha$-element enhancement and more consistent with the  $\alpha/Fe=0.3$ model. This is the well-known  $\alpha$-element enhancement for luminous elliptical galaxies \citep{1976ApJ...206..370O,1989PhDT.......149P,1993MNRAS.262..650D,1995ASPC...86..203W,1999PASP..111..919H,1999MNRAS.306..607J}. 

The solid contours and the dashed contours show the distribution of E+A and the star-forming galaxies with the $H\delta$~EW~$>$~5.0\AA, respectively. 
 First, E+A galaxies are much younger than luminous elliptical galaxies, being consistent with the post-starburst interpretation of E+A galaxies. 
 Interestingly, the peak of the E+A distribution is consistent with $\alpha/Fe=0.3$, suggesting that E+A galaxies have similar amount of  $\alpha$-element enhancement with luminous elliptical galaxies, although the distribution is well extended to  $\alpha/Fe=0.0$. Note that the $Mg_{b}$ vs $Fe(5270)$ plane is not suitable to investigate the total metallicity since it is degenerated with the galaxy age.
Compared to the E+A  galaxies, star-forming galaxies have younger age, and smaller  $\alpha/Fe$ ratio. 

\begin{figure}
\begin{center}
\includegraphics[scale=0.6]{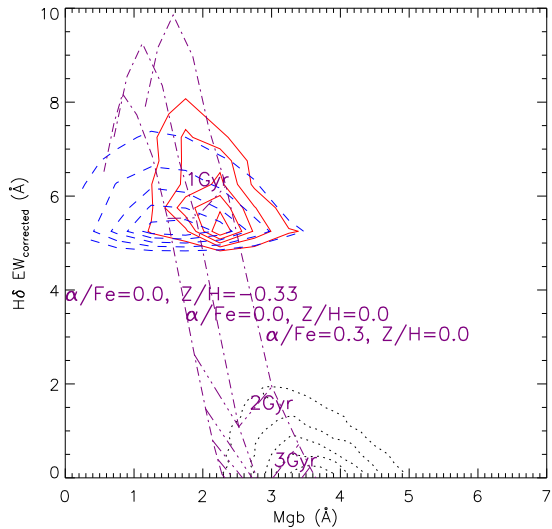}
\end{center}
\caption{
Equivalent width of H$\delta$ is plotted against Mg$_b$. 
 The countour levels are at  16,33,49,66, and 83 percentiles (the
 levels are common to all figures hereafter).
E+A galaxies have slightly higher Mg abundance than the star-forming galaxies.  Both E+A galaxies and elliptical galaxies show $\alpha$-element enhancement compared with the models. H$\delta$ EWs are corrected for emission filling when H$\alpha$ and H$\beta$ emission are present. 
}\label{fig:Hd_Mgb}
\end{figure}

\begin{figure}
\begin{center}
\includegraphics[scale=0.6]{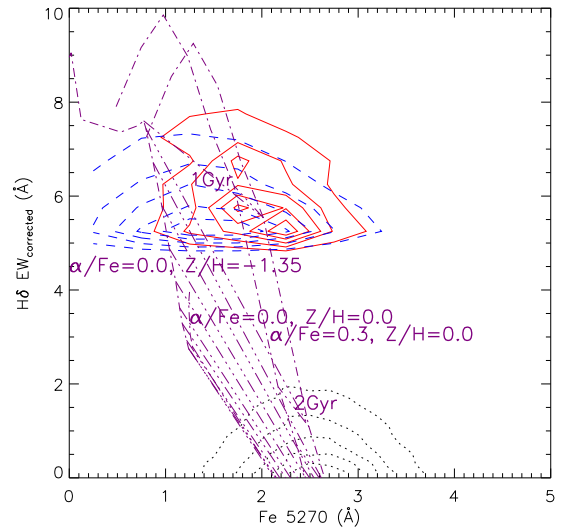}
\end{center}
\caption{
Equivalent width of H$\delta$ is plotted against Fe5270. E+A galaxies have slightly higher Fe abundance than the star-forming galaxies.  Both E+A galaxies and elliptical galaxies show $\alpha$-element enhancement compared with the models. H$\delta$ EWs are corrected for emission filling when H$\alpha$ and H$\beta$ emission are present. 
}\label{fig:Hd_Fe}
\end{figure}

To clarify these points, we plot H$\delta$ EW against $Mg_b$ in Fig. \ref{fig:Hd_Mgb} and against Fe5270 in Fig. \ref{fig:Hd_Fe}. 
 We use the H$\delta$ line here because the line is isolated from other emission and absorption lines, as well as strong continuum features in the galaxy spectrum (e.g.,
D4000). Furthermore, the higher order Balmer lines (H$\gamma$ and H$\beta$)
can suffer from significant emission--filling \citep{Osterbrock}, while the lower order lines (H$\epsilon$ and H$\zeta$) are low signal--to--noise in the spectra.
In both figures, E+A galaxies are better matched with the $\alpha$-enhanced model ($\alpha/Fe=0.3, Z/H$=0.0, the dash-dotted line) with a similar $\alpha$-element enhancement with the passive galaxies in the dotted contours.
 In contrast, the star-forming galaxies are better described with the models with smaller metallicity and no $\alpha$-element enhancement ($Z/H$=-0.33 and -1.35 in Figs.  \ref{fig:Hd_Mgb} and \ref{fig:Hd_Fe}).

\subsection{$\alpha$-element enhancement}

\begin{figure}
\begin{center}
\includegraphics[scale=0.6]{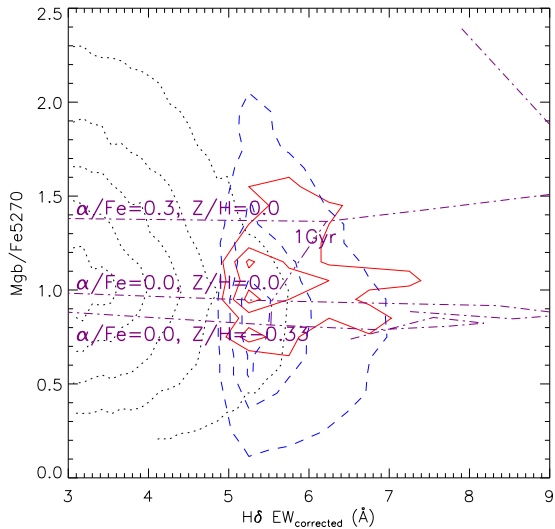}
\end{center}
\caption{
The ratio of Mgb to Fe5270 is plotted against equivalent width of H$\delta$. E+A, star-forming galaxies, and the passive  galaxies are in the solid, dashed, and dotted contours, respectively. The models with  ($\alpha/Fe, Z/H$)=(0.3,0.0), (0.0,0.0) and (0.0,-0.33) are in the dash-dotted lines.
E+A galaxies have slightly higher Fe abundance than the star-forming galaxies.  Both E+A galaxies and elliptical galaxies show $\alpha$-element enhancement compared with the models. H$\delta$ EWs are corrected for emission filling when H$\alpha$ and H$\beta$ emission are present. 
}\label{fig:alpha_hd}
\end{figure}

\begin{figure}
\begin{center}
\includegraphics[scale=0.6]{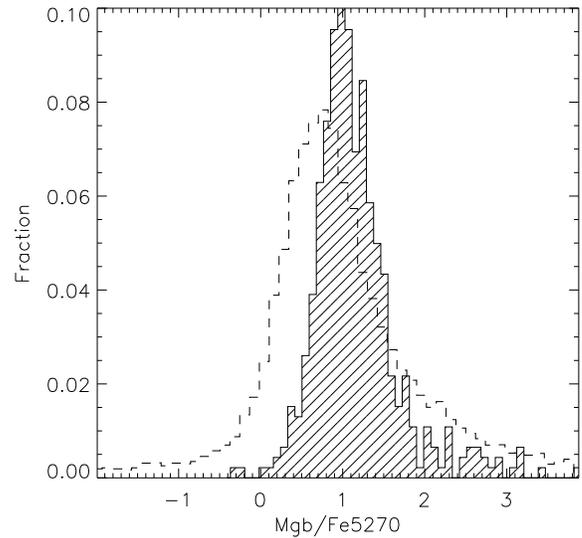}
\end{center}
\caption{
The histogram of the ratio of Mgb to Fe5270 is shown for E+A galaxies in the solid, hashed region, and for star-forming galaxies with the dashed lines. A Kolomogorov-Smirnov test shows these two distributions are significantly different with more than 99.999\% confidence level. 
}\label{fig:histogram}
\end{figure}

To clarify the $\alpha$-element enhancement in E+A galaxies, we plot Mg$_b$/Fe5270 ratio against H$\delta$ EW in Fig. \ref{fig:alpha_hd}. We notice that the  Mg$_b$/Fe5270 distribution of star-forming galaxies are more extended to lower values than that of E+A galaxies. The trend is clearer in Fig. \ref{fig:histogram} where we plot histograms of  Mg$_b$/Fe5270 distributions of E+A galaxies (the hashed region) and star-forming galaxies (the dashed lines). The histogram of E+A galaxies is shifted toward higher values of  Mg$_b$/Fe5270 (i.e., $\alpha$-enhanced). According to the Kolomogorov-Smirnov test, these two distributions are different with more than 99.999\% significance.

\subsection{Metallicity}

\begin{figure}
\begin{center}
\includegraphics[scale=0.6]{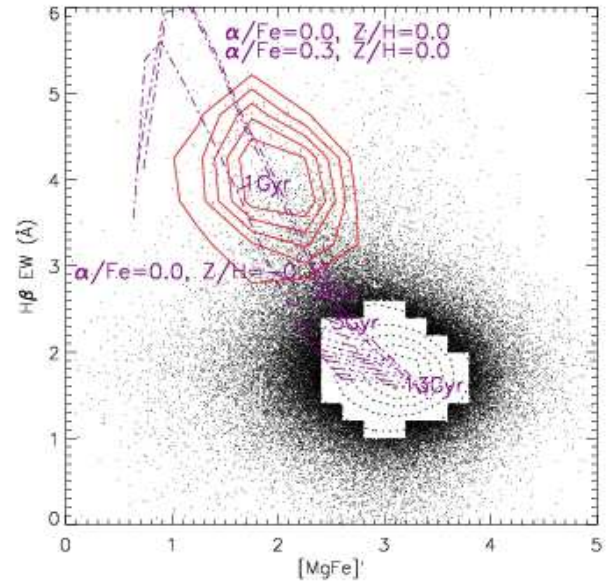}
\end{center}
\caption{H$\beta$ EW is plotted against [MgFe]' index. 
E+A galaxies and the passive galaxies are in the solid and dotted lines, respectively. The models with the dash-dot lines are the models of \citet{2003MNRAS.339..897T} with ($\alpha/Fe, Z/H$)=(0.3,0.0), (0.0,0.0) and (0.0,-0.33).
}\label{fig:hb_mgfe}
\end{figure}

In Fig.\ref{fig:hb_mgfe}, we plot H$\beta$ EW against [MgFe]' index, which is defined by \citet{2003MNRAS.339..897T} as [MgFe]'$\equiv$ $\sqrt{Mg_{b}(0.72\times Fe5270 + 0.28 \times Fe5335})$. The  [MgFe]'  index is independent of $\alpha /Fe$, and hence, a better tracer of total metallicity than the original [MgFe] index suggested by \citet{1993PhDT.......172G}. 
Indeed, two models only different in $\alpha/Fe$ ratio (the dash-dot lines) are almost degenerate in Fig.\ref{fig:hb_mgfe}.
 An advantage in using H$\beta$ line over  H$\gamma$ or H$\delta$ is that  H$\beta$ shows little sensitivity to element ratio variations as well as to total metallicity, because of the lack of strong metallic lines in the wavelength range of the index definition \citep{2005A&A...438..685K}.
 In  Fig.\ref{fig:hb_mgfe}, both E+A galaxies and the passive galaxies are more consistent with the ($\alpha/Fe, Z/H$)=(0.3,0.0) model, with E+A galaxies having younger age. It is suggested that E+A galaxies are as metal rich as present day elliptical galaxies.
 We note that $Z/H=0.0$ is the solar metallicity and is somewhat lower than that of giant metal rich elliptical galaxies. Part of the reason for this is that we include lower mass galaxies in the passive galaxy sample (the dotted lines), and also please note that galaxies are widely distributed around the model \citep[see ][ for a Z/H distribution of local galaxies]{1998A&A...333..419T,2003MNRAS.343..279T}.  In any case, the comparisons within our samples are valid and they still show that E+A galaxies are as metal rich as present day elliptical galaxies.

\begin{figure}
\begin{center}
\includegraphics[scale=0.6]{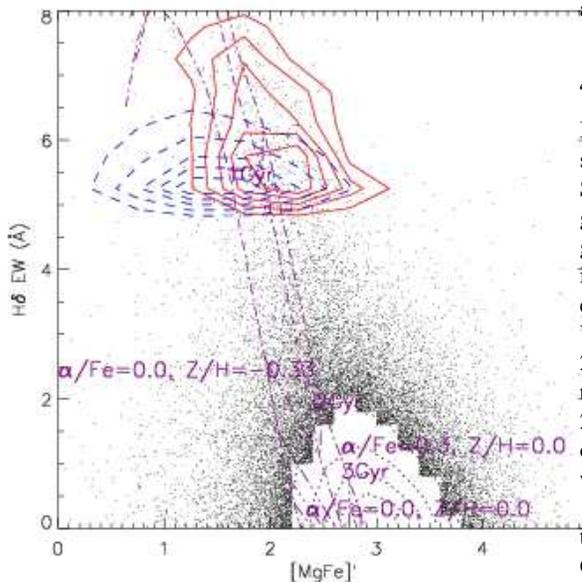}
\end{center}
\caption{H$\delta$ EW is plotted against [MgFe]' index. 
E+A galaxies and the passive galaxies are in the solid and dotted lines, respectively. The models with the dash-dot lines are the models of \citet{2003MNRAS.339..897T} with ($\alpha/Fe, Z/H$)=(0.3,0.0), (0.0,0.0) and (0.0,-0.33).
}\label{fig:hd_mgfe}
\end{figure}

 Although H$\beta$ is less sensitive to metallicity variation, it suffers from the emission filling if the nebular emission is present at the galaxy. One way to reduce emission contamination is to use higher order Balmer lines such as H$\delta$ since they are less affected by nebular emission \citep{Osterbrock}.
 In Fig. \ref{fig:hd_mgfe}, we then plot  H$\delta$ EW against [MgFe]' index. Again both E+A galaxies in the solid contours and the passive galaxies in the dotted contours show good agreement with the ($\alpha/Fe, Z/H$)=(0.3,0.0) model in the dash-dotted line. On the other hand, the star-forming galaxies in the dashed contours agree with the metal poorer model with  ($\alpha/Fe, Z/H$)=(0.0,-0.33).

\section{Discussion}

\subsection{Age-Metallicity Degeneracy}

In the section  \ref{results}, we have measured and interpreted several Lick line indices. Historically, one of the great obstacles in interpreting the galaxy spectra has been the age/metallicity degeneracy in old stellar populations. As pointed out by \citet{1994ApJS...94..687W}, the spectra of an old ($>$2 gigayear) stellar population looks almost identical when the age is doubled and total metallicity reduced by a factor of 3 at the same time.  We partially broke this age-metallicity degeneracy by plotting age-sensitive indices such as H$\delta$ and H$\beta$ against metallicity-sensitive indices such as Mg and Fe lines. Usefulness of this approach has been demonstrated by many authors \citep{1993PhDT.......172G,1995ApJ...448..119F,1998MNRAS.295L..29K,1998PhDT........35M,1999MNRAS.306..607J,2000MNRAS.315..184K}.
Since E+A galaxies have such strong Balmer absorption lines (H$\delta$ EW $>$5\AA), it is rather obvious that E+A galaxies have much younger age ($\sim$1 gigayear) than the comparison sample of elliptical galaxies. It is,  however, important to keep it in mind that the H$\delta$ has $d$log(age)/$d$log($Z$)=0.9 (the age change needed to balance a metallicity change so that the index remains constant)  and that   H$\beta$  has $d$log(age)/$d$log($Z$)=0.6 \citep[][see their Table 7 and Fig.6]{1997ApJS..111..377W} to fully understand the stellar populations of E+A galaxies.
 
%
%

\subsection{$\alpha$-element abundance}
In Section \ref{fig:alpha_hd}, we found that E+A galaxies have high $\alpha$-element abundance ratio, and intepreted that it is another evidence that E+A galaxies are in the post-starburst phase. However, it is important to keep in mind that this is under an assumption that the initial mass function (IMF) is universal. High abundance ratio of $\alpha$ elements can be obtained when the chemical enrichment is driven predominantly by the ejecta of SN II (producer of $\alpha$ elements), which could happen under the following two occasions: (i) when the galaxy truncate its starburst and is in the post-starburst phase; (ii) the fraction of high-mass stars is high, i.e., the top heavy initial mass function \citep{1992ApJ...398...69W}. 
 For E+A galaxies, we not only have the information on the high $\alpha$-elements ratio, but also we know that E+A galaxies have extremely strong Balmer absorption lines (H$\delta$ EW $>$5\AA). Therefore, the IMF has to be not just top heavy, but also need to be A-type star peaked. It is possible, but there is no strong reason that  E+A galaxies have an A-star  peaked IMF. Therefore, it is more straightforward to interpret that our findings of high $\alpha$-elements abundance ratio strengthened the post-starburst interpretation of E+A galaxies.

\subsection{Age of E+A galaxies}

Although  we focused on the metallicity and the $\alpha$-element abundance ratio of E+A galaxies in Section \ref{results}, the models in Figs. \ref{fig:FeMg}-\ref{fig:hd_mgfe} also show galaxy's age. Being consistent with our intension to select post-starburst galaxies, the age distribution of E+A galaxies is peaked at slightly younger than 1 gigayear in Figs. \ref{fig:FeMg}-\ref{fig:hd_mgfe}, which is consistent with the previous picture of E+A galaxies (e.g., Possianti et al. 1999). However, it is important to note that this age estimation is based on the single stellar population model, which experienced the burst of star formation at the zero age, followed by the truncation of  star formation. In reality, the truncation of star formation needs a short, but finite length of time scale, and therefore, it is more practical approach to use a model with variable truncation time scale of star formation (e.g., 0.25-1 gigayear). An initial attempt can be found in \citet{2004A&A...427..125G,2005MNRAS.359.1557Y}.


\subsection{Are E+A galaxies progenitors of elliptical galaxies?}

 In Section \ref{results}, we have collected evidence that E+A galaxies have similar $\alpha$/Fe ratio and metallicity with luminous elliptical galaxies. Since the only difference in our plots between these populations of galaxies is the age, E+A galaxies will surely evolve into elliptical galaxies in a few gigayear. In addition, there has been accumulating evidence that the morphological appearance of E+A galaxies is either early-type or with dynamically disturbed signs \citep{2005MNRAS.357..937G,2005MNRAS.359.1557Y,2006ApJ...642..152Y,2006AJ....131.2050Y}. The surface brightness of E+A galaxies is also inbetween that of elliptical galaxies and star-forming galaxies \citep{2004ApJ...602..190Q}.  These results suggest that E+A galaxies also look like one of  the progenitors of  elliptical galaxies in terms of morphology as well. 

It has been reported, however, that there is a huge difference in the number density between E+A galaxies and elliptical galaxies; \citet{2005MNRAS.357..937G} reported that E+A galaxies are only 0.2\% of galaxies in a volume-limited sample. 
 Assuming that the lifetime of the E+A phase is $\sim$1 gigayear based on the lifetime of A-type stars, only 2-3\% of galaxies have been through the E+A phase in the Universe of 13.7 gigayears of age, if the fraction of the E+A galaxies has been constant at $\sim$0.2 \%. This, however, is a too simple assumption. It has been known that cosmic star formation density was much higher at $z>1$ \citep{1996MNRAS.283.1388M}. According to \citet{2006ApJ...642...48L}, the fractions of the post-starburst galaxies was larger by a factor of 2 at $z\sim 1$. If we loosen our criteria to select post-starburst galaxies to H$\delta$\_EW $>$3\AA, the fraction increase by a factor of $\sim$3 \citep{2003PASJ...55..771G}. Therefore, an estimate on the high side yield $\sim$15\% of galaxies may have been through the E+A phase. However, this is still a small fraction of galaxies compared to the total number of galaxies and suggests that not all the elliptical galaxies have not been through the E+A phase. 
 Therefore, it is likely that E+A galaxies are only one of the multiple progenitors of elliptical galaxies.
 If star-forming galaxies decrease their star formation rate more slowly, for example, exponentially with $\tau\sim$1 gigayear, the galaxies do not experience the E+A phase before they become passive galaxies \citep{2003PhDT.........2G}. 
It is important to investigate what special physical mechanism is needed for a galaxy to become an E+A galaxy. 

 Specifically on the high redshift cluster environments, there is a
 conflicting  report on the fraction of E+A galaxies: \citet{2003ApJ...599..865T}  found 7-13\% of E+A galaxies in three high redshift clusters at z=0.33,0.58, and 0.83, claiming that $>30$\% of E+S0 members may have undergone the E+A phase if the effects of E+A downsizing and increasing E+A fraction as a function of redshift are considered. In their search for field E+A galaxies among 800 spectra, \citet{2004ApJ...609..683T}   measured the E+A fraction at 0.3 $< z <$ 1 to be 2.7\% $\pm$ 1.1\%, a value lower than that in galaxy clusters at comparable redshifts.
 Based on 1823 galaxies in the 15 X-ray luminous clusters at 0.18 $< z <$ 0.55, on the contrary, \citet{1999ApJ...527...54B} found that the fraction of K+A galaxies is 1.5\% $\pm$ 0.8\% in the cluster and 1.2\% $\pm$ 0.8\% in the field, i.e., they found no difference. 

 Obviously these two papers are inconsistent to each other.  Both of these work did not use H$\alpha$ line in the selection criteria, and thus, could suffer from the up to 52\% of the contamination \citep{2003PASJ...55..771G}. More statistically significant analysis based on a larger sample is needed to shed light on the galaxy evolution in the high redshift cluster environment. 
 In the low redshift cluster environment, \citep{2003PASJ...55..757G} identified that passive spiral galaxies, i.e.,  galaxies with no star formation but with spiral morphology, preferentially reside in the cluster infalling regions. These passive spiral galaxies may be a more important smoking-gun in galaxy evolution in cluster environment.

\section{Conclusions}

We have constructed one of the largest sample of 451 E+A galaxies using the SDSS DR4. Using the sample we have investigated total metallicity and $\alpha$-element enhancement of E+A galaxies. Our findings are as follows:

\begin{enumerate}
 \item Compared to the star-forming galaxies with similar amount of Balmer absorption, E+A galaxies have larger amount of metallicity and $\alpha$-element enhancement. These results are consistent with the hypothesis that the star formation in E+A galaxies were truncated to enhance the  $\alpha$-element abundance in E+A galaxies. This  provides us with additional evidence that E+A galaxies selected to have no H$\alpha$ emission are more likely to be truly in the post-starburst phase rather than dusty-starforming galaxies.
 \item Compared to the present-day elliptical galaxies, E+A galaxies have similar amount of metallicity and $\alpha$-element enhancement. These results suggest that in terms of abundance diagnosis, E+A galaxies are consistent to be one of the progenitors of the present-day elliptical galaxies, although the difference in the spatial number density needs to be carefully considered.
 \item Compared with the models of \citet{2003MNRAS.339..897T} which can handle the difference in the $\alpha$/Fe ratio at all metallicities, E+A galaxies are well-explained with ($\alpha/Fe, Z/H$)=(0.3,0.0) and age of $<$1 gigayear.
\end{enumerate}

\section*{Acknowledgments}

We thank Dr. M.Yagi, and C.Yamauchi for useful suggestions.
We thank the anonymous referee for many insightful comments, which significantly improved the paper.
We thank V.Ivanov, I.Saviane and the ESO Office Santiago for their hospitality during our visit.


 The research was financially supported by the Sasakawa Scientific Research Grant from The Japan Science Society.

   
    Funding for the creation and distribution of the SDSS Archive has been provided by the Alfred P. Sloan Foundation, the Participating Institutions, the National Aeronautics and Space Administration, the National Science Foundation, the U.S. Department of Energy, the Japanese Monbukagakusho, and the Max Planck Society. The SDSS Web site is http://www.sdss.org/.

    The SDSS is managed by the Astrophysical Research Consortium (ARC) for the Participating Institutions. The Participating Institutions are The University of Chicago, Fermilab, the Institute for Advanced Study, the Japan Participation Group, The Johns Hopkins University, Los Alamos National Laboratory, the Max-Planck-Institute for Astronomy (MPIA), the Max-Planck-Institute for Astrophysics (MPA), New Mexico State University, University of Pittsburgh, Princeton University, the United States Naval Observatory, and the University of Washington.



\label{lastpage}

\end{document}